\documentclass{pasa}%
 
\usepackage{graphicx}
\usepackage{natbib,epsfig,amsmath,rotating,color,upgreek} 
\usepackage{times}

\def\kms{km ${\rm s}^{-1}$}

\def\Mo{M$_\odot$}

\def\ccm {$\hbox{{\rm cm}}^{-3}$}    
\def\scm  {$\hbox{{\rm cm}}^{-2}$}    
\def \AL {$\alpha $}     
\def \HI {H\,{\small I}}   

\def\MOLH {\hbox{${\rm H}_2$}}  
\def\dg{$^{\circ}$}
\def\lapp{\ifmmode\stackrel{<}{_{\sim}}\else$\stackrel{<}{_{\sim}}$\fi}
\def\gapp{\ifmmode\stackrel{>}{_{\sim}}\else$\stackrel{>}{_{\sim}}$\fi}
\title[Cold gas fraction and the star formation history]{Evolution of the cold gas fraction and  the star formation history: Prospects with current and future radio facilities}
\author[S. J. Curran]{S. J. Curran\thanks{Stephen.Curran@vuw.ac.nz}
\affil{School of Chemical and Physical Sciences, Victoria University of Wellington, PO Box 600, Wellington 6140, New Zealand}
}

\jid{PASA}
\doi{10.1017/pas.\the\year.xxx}
\jyear{\the\year}

\usepackage{aas_macros} 
 
\begin{document}

\begin{frontmatter} 
\maketitle

\begin{abstract}
  It has recently been shown that the abundance of  cold neutral gas may follow a similar evolution as the star
  formation history. This is physically motivated, since stars form out of this component of the neutral gas and if the
  case, would resolve the longstanding issue that there is a clear disparity between the total abundance of neutral gas
  and star forming activity over the history of the Universe. Radio-band 21-cm absorption traces the cold gas and
  comparison with the Lyman-\AL\ absorption, which traces all of the gas, provides a measure of the cold gas fraction,
  or the spin temperature, $T_{\rm spin}$.  The recent study has shown that the spin temperature (degenerate with the
  ratio of the absorber/emitter extent) appears to be anti-correlated with the star formation density, $\psi_{*}$, with
  $1/T_{\rm spin}$ undergoing a similar steep evolution as $\psi_{*}$ over redshifts of $0 \lapp z \lapp 3$, whereas the
  total neutral hydrogen exhibits little evolution. Above $z\sim3$, where $\psi_{*}$ shows a steep decline with
  redshift, there is insufficient 21-cm data to determine whether $1/T_{\rm spin}$ continues to follow
  $\psi_{*}$. Knowing this is paramount in ascertaining whether the cold neutral gas does trace the star formation over
  the Universe's history.  We explore the feasibility of resolving this with 21-cm observations of the largest
  contemporary sample of reliable damped Lyman-\AL\ absorption systems and conclude that, while today's largest radio
  interferometers can reach the required sensitivity at $z\lapp3.5$, the Square Kilometre Array is required to probe to
  higher redshifts.
\end{abstract}

\end{frontmatter}

\section{INTRODUCTION }
\label{intro}

Neutral hydrogen (\HI) provides the raw material for the formation of stars and so the abundance of this at various
redshifts should closely trace the star forming activity over the history of the Universe. Specifically, the star
formation rate density, $\psi_{*}$, exhibits a steep climb from $z=0$ before peaking at $z\sim2.5$, followed by a steep
decline at higher redshifts \citep{hb06,bbg+13,ssb+13,lbz+14,md14,zjd+14}. However, the mass density of the neutral gas
exhibits very little redshift evolution: Over  $0\lapp z\lapp0.5$, where \HI\ 21-cm can be detected in emission
(e.g. \citealt{fgv+16}), the mass density is $\Omega_{\rm HI} \approx0.5\times10^{-3}$ 
\citep{zvb+05,lcb+07,bra12,dsmb13,rzb+13,hsf+15}. At higher redshift, where the \HI\ column density is obtained from
either space ($z\lapp1.7$) or ground ($z\gapp1.7$) based observations of damped Lyman-$\alpha$ systems (DLAs, where
$N_{\rm HI}\ge2\times10^{20}$ \scm)\footnote{These gas-rich galaxies, detected through the Lyman-\AL\ absorption of a
  background continuum source, may account or at least 80\% of the neutral gas mass density in the Universe
  \citep{phw05}.}, the mass density rises to $\Omega_{\rm HI} \approx1\times10^{-3}$, remaining constant up to at least
$z\sim5$ \citep{rt00,ph04,rtn05,cur09a,pw09,npc+12,cmp+15,npr+16}.  Thus, there is a clear disparity between the star
formation density and the neutral gas available to fuel it.

\begin{figure}
\centering \includegraphics[angle=-90,scale=0.55]{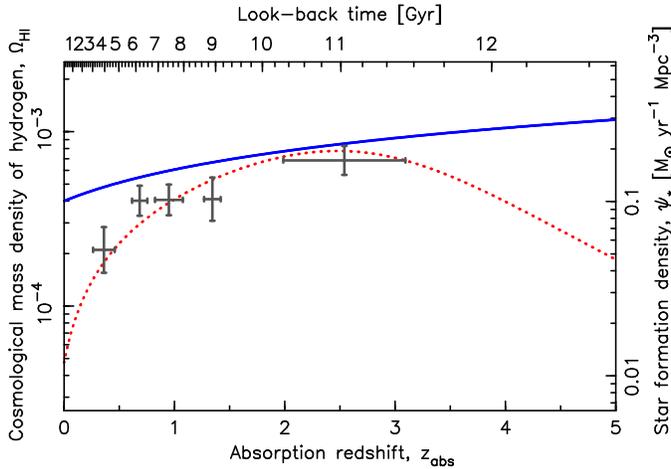}
\caption{The cosmological mass density of neutral hydrogen (solid trace, \citealt{cmp+17}) and the star formation density (dotted trace,
  \citealt{hb06}) versus redshift.  The error bars show the binned ($n=50$ per bin) $\pm1\sigma$ values of $(1/T_{\rm spin})(d_{\rm abs}/d_{\rm QSO})^2$, normalised on the ordinate by $500$~\Mo\ yr$^{-1}$ Mpc $^{-3}$ K \citep{cur17a}.}
\label{SFR-today}
\end{figure}

The Lyman-$\alpha$ ($\lambda =1215.67$~\AA ) transition traces all of the neutral gas, specifically both the cold (CNM,
where $T\sim150$~K and $n\sim10$~\ccm) and warm neutral media (WNM, where $T\sim10\,000$~K and $n\sim0.2$~\ccm,
\citealt{fgh69,whm+95}).  Given that only the gas clouds which are cool enough to collapse under their own gravity are
expected to initiate star formation, we may only expect the cool component, the CNM, to follow the star formation
history. Radio-band 21-cm absorption traces this component of the hydrogen and the comparison of this with the total
column density provides a temperature measure of the gas. Using this method, \citet{cur17}
showed that the spin temperature and the star formation density may be inversely related (Fig.~\ref{SFR-today}).
As seen from the figure, however, while $1/T_{\rm spin}$ (degenerate with the ratio of the absorber and emitter
extents, $d_{\rm abs}/d_{\rm QSO}$), does follow a similar increase as $\psi_{*}$ up to the $z\sim2.5$
peak, what happens beyond this is as yet unknown. In this paper, we explore the feasibility of obtaining the required
data with current state-of-the-art radio telescopes, in addition to exploring the prospects with the {\em Square Kilometre Array}
(SKA).

\section{The sample}
\label{cds}

From Fig. \ref{SFR-today} it is clear that higher redshift ($z_{\rm abs}\gapp3$) data are required in order to determine
whether the spin temperature, degenerate with the ratio of the absorber/emitter extent, is anti-correlated with the star
formation density, or whether it just (coincidentally) increases by the same approximate factor out to the peak
$\psi_{*}$. Since we wish to find the typical sensitivity limits attainable by current and future facilities, we require
a large readily available sample of DLAs. Such a sample is provided by the spectra of quasi-stellar objects (QSOs) in
the {\em Sloan Digital Sky Survey} (SDSS), which provides the largest catalogue of DLAs \citep{ph04,phw05, npls09}, with
the SDSS-III DR9 DLA \citep{npc+12} being the most recently published catalogue of reliable absorbers.\footnote{The DLAs
  in the {\em Baryon Oscillation Spectroscopic Survey} (BOSS) of the SDSS-III \citep{ppa+14} are, as yet, unconfirmed.}
This contains 12\,081 DLAs and sub-DLAs (all with $N_{\rm HI}\ge1\times10^{20}$ \scm), over redshifts of $1.951\leq
z_{\rm abs} \leq 5.343$.

\begin{figure}
\centering \includegraphics[angle=-90,scale=0.52]{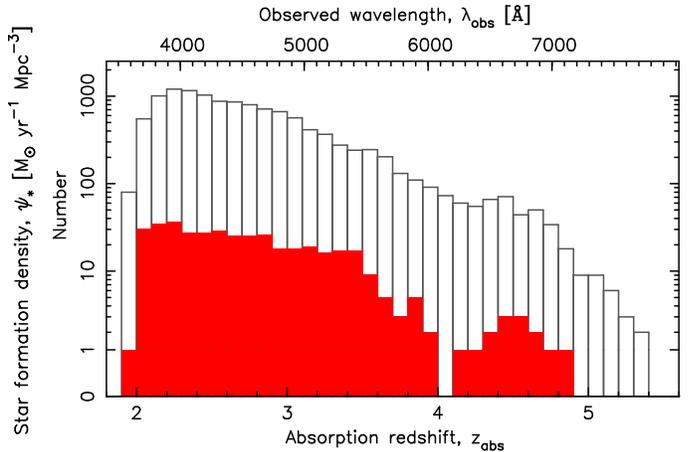}
\caption{The redshift distribution of the DR9 DLAs (unfilled histogram) overlaid with those occulting a known radio source
(filled histogram).}
\label{DR9-histo}
\end{figure}

We match each QSO to the nearest radio source (within 10 arc-sec) in each of the {\em NRAO VLA Sky Survey}
(NVSS, \citealt{ccg+98}), the Very Large Array's {\em Faint Images of the Radio Sky at Twenty-Centimeters} (FIRST,
\citealt{wbhg97}) and the {\em Sydney University Molonglo Sky Survey} (SUMSS, \citealt{mmb+03}). This yields 336 radio-loud
sight-lines, containing a total of 414 absorbers (Fig. \ref{DR9-histo}).
For each of the large radio interferometers, the {\em Murchison Widefield Array} (MWA), the {\em Low-Frequency Array}
(LOFAR), the {\em Giant Metrewave Radio Telescope} (GMRT), the {\em Very Large Array} (VLA), as well as the forthcoming
SKA, we determine which of the DLAs occulting a radio-loud source would have the 21-cm transition redshifted into an
available band.  From an estimate of the flux density, we then determine the best expected sensitivity of the instrument
and whether this is sufficient to detect the putative high redshift downturn in $1/T_{\rm spin}$
(Fig. \ref{SFR-today}). Note that we do not include large single dish instruments (e.g. the {\em Green Bank Telescope}, GBT) due to the restrictions in mitigating
the radio frequency interference (RFI) at these frequencies (e.g. \citealt{caw+16}).

\section{Analysis}
\subsection{Attainable limits}
\label{sec:sens}

From the radiometer equation (e.g. \citealt{rw00}), 
the  r.m.s. noise level obtained after an integration time of $t_{\rm int}$ is given by
\begin{equation}
\sigma_{\rm rms} = \frac{2k_{\rm B}T_{\rm sys}}{\epsilon_{\rm c} A_{\rm eff}}\frac{1}{\sqrt{n_{\rm p}\Delta\nu t_{\rm int}}},
\label{eq:rad}
\end{equation}
where $k_{\rm B}$ is the Boltzmann constant, $T_{\rm sys}$ is the system temperature, $\epsilon_{\rm c}$ the correlator
  efficiency,
$A_{\rm eff}$ the effective collecting area of the telescope, 
$n_{\rm p}$ the number of polarisations and  $\Delta\nu$ is the channel bandwidth.
This is related to the velocity resolution, $\Delta v$, via $\Delta\nu =  (\Delta v/c)\nu_{\rm obs}$, where $c$ is the
speed of light and $\nu_{\rm obs}$ is the observed frequency.

The observed optical depth, $\tau_{\rm obs}$, of the absorption is given by the ratio of the line depth, $\Delta S$, to
the observed background flux, $S$, and is related to the intrinsic optical depth, $\tau$,  via
\begin{equation}
\tau \equiv-\ln\left(1-\frac{\tau_{\rm obs}}{f}\right) \approx  \frac{\tau_{\rm obs}}{f},
\label{tau_obs}
\end{equation}
in the optically thin regime ($\Delta S/S\lapp0.3$), where the covering factor, $f$, is the fraction of $S$
intercepted by the absorber. In the case of a $3\sigma$ upper limit, the optical depth is thus given 
by $\tau_{\rm obs}\lapp 3\sigma_{\rm rms}/S$.

Using the flux density (see Sect. \ref{sec:rf}), we can therefore estimate the upper limits to the observed velocity
integrated optical depth. Inserting $\int\!\tau_{\rm obs}\,dv \leq
(3\sigma_{\rm rms}/S)\Delta v$ per $\Delta v$ channel (see \citealt{cur12}) into   $N_{\rm HI}
=1.823\times10^{18}\,T_{\rm spin}\int\!\tau\,dv$ \citep{wb75}, gives 
\begin{equation}
\frac{T_{\rm  spin}}{f} \gapp\frac{N_{\rm HI}}{1.823\times10^{18}  (3\sigma_{\rm rms}/S)\Delta v}.
\label{eq:spin}
\end{equation}
Providing that the Lyman-\AL\ and 21-cm absorption arise along the same sight-line, Equ. \ref{eq:spin} yields the limit
to the spin temperature of the gas attainable by each instrument.  $T_{\rm spin}$ is a measure of the excitation of the
gas by 21-cm absorption \citep{pf56}, excitation above ground state by Lyman-\AL\ absorption \citep{fie59} and
collisional excitation \citep{be69}.  This provides a thermometer, allowing a measure of the fraction of the CNM -- the cool
gas in which star formation occurs. The spin temperature is, however, degenerate with the covering factor (Equ.~\ref{eq:spin}),
whose value requires knowledge of the relative extents of the absorber--frame 1420~MHz absorption and emission
cross-sections ($d_{\rm abs}/d_{\rm QSO}$). This can, however, be expressed as
\begin{equation}
f= \left\{   
\begin{array}{l l}
\left(\frac{d_{\rm abs} DA_{\rm QSO}}{d_{\rm QSO} DA_{ \rm abs}}\right)^2 & \text{ if } \theta_{\rm abs} < \theta_{\rm QSO},\\
  1  & \text{ if } \theta_{\rm abs} \geq\theta_{\rm QSO}\\
\end{array}
\right.  
\label{f}
\end{equation}
where 
$DA_{\rm abs}$ and $DA_{\rm QSO}$ are the angular diameter
distances and $\theta_{\rm abs}$ and $\theta_{\rm QSO}$ the angular extents of the absorbers and emitter, respectively
(\citealt{cur12}). 

The angular diameter distance is obtained  from line-of-sight co-moving distance, $DC$, (e.g. \citealt{pea99}), 
\[
DA =  \frac{DC}{z+1}, {\rm ~where~} DC = \frac{c}{H_0}\int_{0}^{z}\frac{dz}{H_{\rm z}/H_0}
\]
in which $c$ is the speed of light, $H_0$ the Hubble constant, $H_{\rm z}$ the Hubble parameter at redshift $z$ and 
\[
\frac{H_{\rm z}}{H_{0}} = \sqrt{\Omega_{\rm m}\,(z+1)^3 + (1-\Omega_{\rm m} - \Omega_{\Lambda})\,(z+1)^2 + \Omega_{\Lambda}}.
\]
For a standard $\Lambda$ cosmology, with $H_{0}=71$~km~s$^{-1}$~Mpc$^{-1}$, $\Omega_{\rm m}=0.27$ and
$\Omega_{\Lambda}=0.73$, this gives a range of possible values of $DA_{\rm abs}/DA_{\rm QSO}$ at $z_{\rm abs}\lapp1.6$,
whereas above this redshift the ratio of angular diameter distances is {\em always} close to unity, i.e. $DA_{\rm abs}/DA_{\rm
  QSO}\sim1, ~\forall z_{\rm abs}\gapp1.6$.  If left uncorrected, this will introduce a bias to the covering factors between
low and high redshift DLAs \citep{cw06}.  Correcting for this, by inserting Equ. \ref{f} into Equ. \ref{eq:spin}, gives
\begin{equation}
T_{\rm  spin}\left(\frac{d_{\rm QSO}}{d_{\rm abs}}\right)^2 \gapp\frac{N_{\rm HI}(DA_{\rm QSO}/DA_{\rm abs})^2}{1.823\times10^{18}  (3\sigma_{\rm rms}/S)\Delta v},
\label{eq:f}
\end{equation}
for $f<1$, which is expected for the high redshift absorbers \citep{cur17}.  

Binning the current 21-cm absorption data, the spin temperature degenerate with the ratio of the emitter--absorber
extent, $(1/T_{\rm spin})(d_{\rm abs}/d_{\rm QSO})^2$, traces the star formation density up to the redshift limit of the
data (Fig.~\ref{SFR-today}) and reproduces the observed CNM fractions in DLAs for a mean $\left<d_{\rm QSO}/d_{\rm
    abs}\right> \sim4$ \citep{cur17a}.  With Equ. \ref{eq:f}, we can estimate the attainable limit for each radio
illuminated DLA in the SDSS DR9 and whether its addition to the current data could help determine if $(1/T_{\rm
  spin})(d_{\rm abs}/d_{\rm QSO})^2$ follows the $\psi_{*}$ downturn at $z_{\rm abs}\gapp3$.

\subsection{Radio fluxes} 
\label{sec:rf}

\begin{figure}
\centering \includegraphics[angle=-90,scale=0.52]{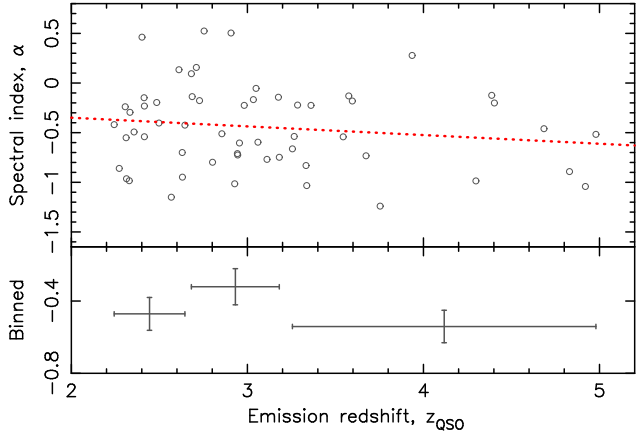}
\caption{The radio-band spectral index versus redshift for the 58 SEDs fit by a power-law, where the dotted line shows
  the least-squares fit.  The bottom panel shows the binned values in equally sized bins, where the horizontal error
  bars show the range of points in the bin and the vertical error bars the error in the mean value.}
\label{SI_z}
\end{figure}

\begin{figure*}
\centering \includegraphics[angle=-90,scale=0.75]{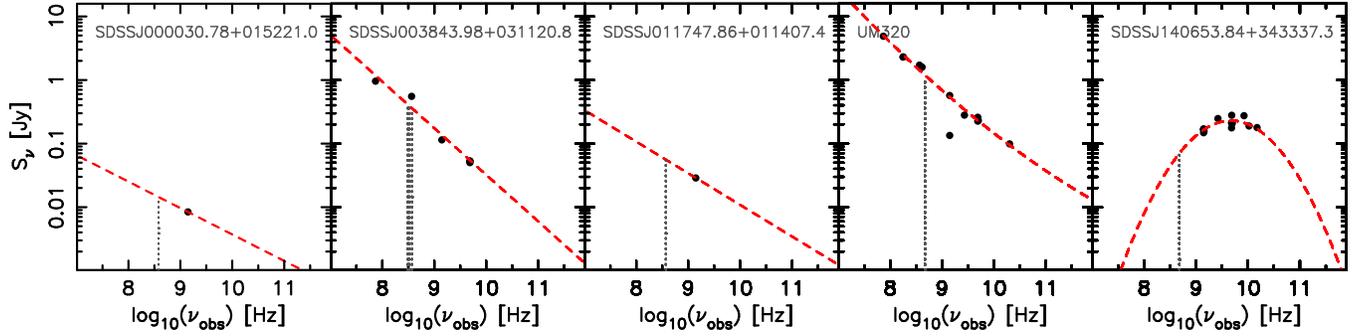}
\caption{Examples of SED fits to the radio photometry of the background sources. The vertical dotted lines show the 
frequency of the 21-cm transition at the absorber redshift (the second source has four intervening absorbers).}
\label{SEDs}
\end{figure*}

\begin{figure}
\centering \includegraphics[angle=-90,scale=0.55]{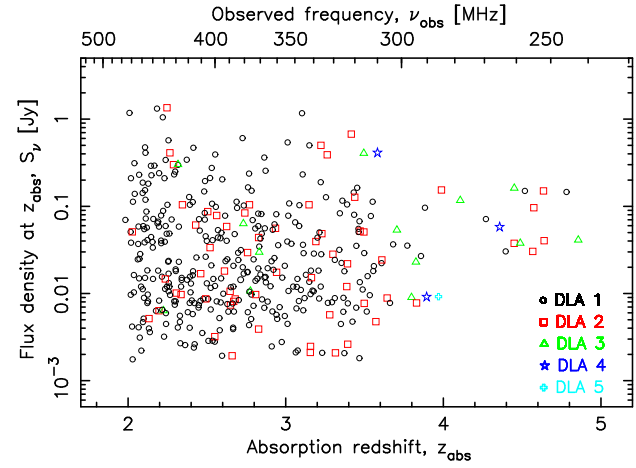}
\caption{The flux density estimated at the absorption redshift.  The colour/shape of the symbol indicates which
  absorber, since some sight-lines have multiple -- 48 sight-lines with a second DLA, ten with a third, two with a fourth and one
  with a fifth. }
\label{flux-z}
\end{figure}

For each of the 336 radio-loud sight-lines (Sect. \ref{cds}), we obtain the radio photometry from NASA/IPAC Extragalactic
Database (NED) 
and estimate the flux density from the background
source at the redshifted 21-cm frequency of each DLA, via a fit in logarithm space:
\begin{itemize}
\item[--] For at least four radio-band photometry points; a second-order polynomial, according to the prescription of \citet{cwsb12},
\item[--]  for two or three radio-band photometry points; a first-order polynomial (a power-law),
\item[--]  for a single radio-band photometry point we assume a spectral index. Values of  $\alpha\approx-1$ are 
typical of $z\gapp3$ radio sources (e.g. \citealt{dvs+02,cwsb12}), although there may be a redshift dependence (\citealt{ak98}).
\end{itemize}
There are 14 sources which are fit by the second-order polynomial and 58 fit by a power-law, leaving 
264 requiring an estimate of the spectral index. In Fig. \ref{SI_z} we show how the spectral indices of the 58 power-law fits vary with redshift.
From this, it is clear that, since the measured flux is generally at a higher frequency than the redshifted 21-cm
absorption (Fig. \ref{SEDs}), assuming too steep a spectral index could significantly overestimate the flux density.
Rather than assuming $\alpha=-1$, we therefore use a redshift dependent spectral index, where $\alpha = -0.087\,z_{\rm
  QSO} -0.175$ (Fig. \ref{SI_z}). The resulting flux density distribution is shown in Fig. \ref{flux-z}.

\subsection{Prospects with current instruments}

\subsubsection{Murchison Widefield Array}

\begin{figure}
\centering \includegraphics[angle=-90,scale=0.55]{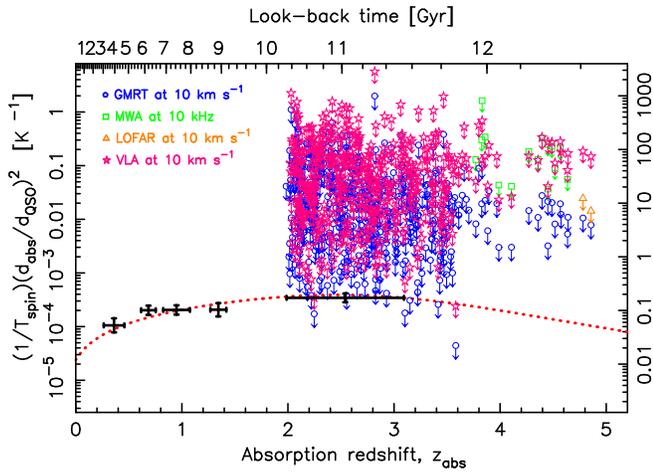}
\caption{As Fig. \ref{SFR-today}, but showing the $3\sigma$ limits in $(1/T_{\rm spin})(d_{\rm abs}/d_{\rm QSO})^2$
  reached after ten hours of integration at a spectral resolution of 10 \kms\ per channel for the current instruments.}
\label{sens-current}
\end{figure}

The MWA (\citealt{tgb+13}) has a coverage of 80--300 MHz, which spans \HI\ 21-cm at redshifts of $z_{\rm abs} > 3.735$.
For declinations of $\delta < 33$\dg, this range gives 15 DLAs illuminated by a radio source observable by the MWA. The
spectrometer is limited to a channel spacing of $\geq10$ kHz, which corresponds to a spectral resolution of $\Delta
v\geq10$~\kms\ at $\nu_{\rm obs}\leq300$~MHz.\footnote{The full-width half maximum of the 21-cm absorption in DLAs has a
  mean value of $\left< {\rm FWHM}\right> = 33$~\kms\ \citep{cur17}.}  At $\nu_{\rm obs} =200$ MHz, $T_{\rm sys} =
195$~K and using all 128 tiles ($A_{\rm eff} = 2534$ m$^2$), gives a theoretical r.m.s. noise level of $\sigma_{\rm rms}
=11$ mJy per 10 kHz channel after $t_{\rm int} = 10$ hours per sight-line (Equ.~\ref{eq:rad}).\footnote{A correlator
  efficiency of $\epsilon_{\rm c}=1$ is assumed \citep{tgb+13}.}  From a search for \HI\ 21-cm absorption within the
hosts of high redshift radio sources\footnote{Proposal ID G0036 (Allison et al.).}, a multiplicative (``fudge'') factor
of $\approx2$ is found, i.e. we should expect an r.m.s. noise level of $\sigma_{\rm rms}\approx22$ mJy per 10 kHz
channel.  Combining this with the estimated flux densities, we obtain limits of $(1/T_{\rm spin})(d_{\rm abs}/d_{\rm
  QSO})^2\gapp0.01$ K$^{-1}$, which are insufficient to confirm the putative downturn in $(1/T_{\rm spin})(d_{\rm
  abs}/d_{\rm QSO})^2$ at $z\gapp3$ (Fig.~\ref{sens-current}).

\subsubsection{Low-Frequency Array}
\label{sec:lof}

\begin{figure}
\centering \includegraphics[angle=-90,scale=0.55]{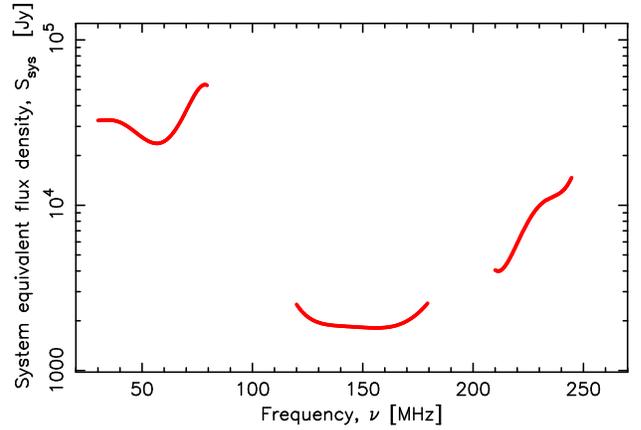}
\caption{Polynomial fits to the system equivalent flux density of LOFAR \citep{vwg+13}.}
\label{LOFAR}
\end{figure}

LOFAR \citep{vwg+13} has a frequency range of  10--250 MHz, which covers the 21-cm transition for 
only two of the DR9 DLAs, with the $\delta > -7$\dg\ declination limit imposing no additional  cut. From the point source
sensitivity (Fig. \ref{LOFAR}), 
$t_{\rm int} = 10$ hours  gives an r.m.s. noise level of $\sigma_{\rm rms} =11-13$ mJy per 10~\kms\ channel,
cf. the theoretical 1.5 mJy using all stations (24 core, 14 remote \& 13 international). A fudge factor of $\lapp10$ in the noise level is expected\footnote{https://support.astron.nl/ImageNoiseCalculator/sens.php},  
giving the values in Fig. \ref{LOFAR}, although spectral line observations suggest
that this should be 3--5  (e.g. \citealt{ovs+14}). Applying a factor of 4 gives $\sigma_{\rm rms} =5.4-6.5$ mJy per 10~\kms\ channel and
limits of $(1/T_{\rm spin})(d_{\rm abs}/d_{\rm QSO})^2\sim0.01$ K$^{-1}$, which are, again, insufficient to detect the
downturn (Fig.~\ref{sens-current}).


\subsubsection{(Upgraded) Giant Metrewave Radio Telescope}

For the GMRT \citep{ana95}, there are two bands available which can detect 21-cm over the redshift range of the DR9 DLAs --- 
the 230--250 MHz  
and 250--500 MHz bands, the latter of which 
is currently being commissioned as part of the upgraded GMRT (uGMRT). Combining these, gives 414 sight-lines towards a
radio-loud source for which 21-cm is redshifted to 230--500 MHz and is at a declination observable with the GMRT
($\delta\gapp-40$\dg).
In order to calculate the telescope sensitivity, we estimate the system temperature by interpolating the values quoted
for the 151, 235 and 325 MHz bands\footnote{http://gmrt.ncra.tifr.res.in/$\sim$astrosupp/obs\_setup/sensitivity.html},
giving $T_{\rm sys} = 0.0175\nu_{\rm obs}^{~~~2} -11.2\nu_{\rm obs} + 1915$, where $\nu_{\rm obs} $ is in MHz, resulting
in $T_{\rm sys} = 106 - 215 $ K for the sample. 
For a $t_{\rm int} = 10$ hour integration using all 30 antennas ($A_{\rm eff} = 26\,508$~m$^2$) and a fudge factor 
of 5 in the time estimate\footnote{No correlator efficiency is quoted for the GMRT, although the integration time is multiplied by a
 factor of $\leq10$, depending upon the observing band
  (http://gmrt.ncra.tifr.res.in/$\sim$astrosupp/obs\_setup/sensitivity.html).}, we reach $\sigma_{\rm rms} =1.3-16$~mJy
per 10~\kms\ channel.

\begin{figure}
\hspace{-1.5em}\centering \includegraphics[angle=-90,scale=0.55]{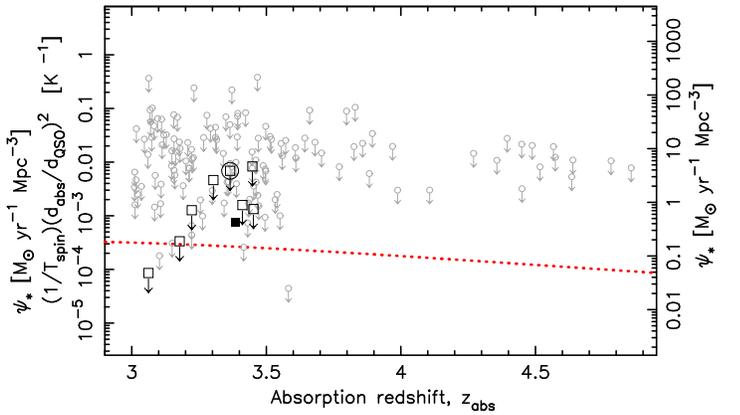}
\caption{A detail of Fig. \ref{sens-current} showing the current $z_{\rm abs} >3$ results (Table~\ref{tab1}), re-sampled
  to a 10 \kms\ spectral resolution and a 10 hour integration, overlain upon the predicted GMRT limits. The filled square
  signifies the 21-cm detection and the circled square the only non-GMRT observation.}
\label{sens_GMRT}
\end{figure}

The radiometer equation (Equ. \ref{eq:rad}) gives the theoretical noise level, which can be significantly lower than the
actual value obtained in the production of an image (e.g. Sect. \ref {sec:lof}).  The fudge factor is intended to
correct for this and, in order to check the validity of the value used, we show actual measured sensitivities
(Table~\ref{tab1}) in comparison to our predicted limits (Fig.~\ref{sens_GMRT}).
\begin{table}
  \caption{The $z_{\rm abs} >3$ DLAs searched for \HI\ 21-cm absorption. $\sigma_{\rm rms}$ [mJy] is the r.m.s. noise level per
    $\Delta v$ [\kms] channel after $t_{\rm int}$ [hours] obtained with the listed telescope. The absorber towards B0201+113 is detected in 21-cm absorption with $\int\tau_{\rm obs}dv = 0.71$ \kms.}
\begin{tabular}{@{}lcccccc@{}}
\hline\hline
IAU name& $z_{\rm abs}$  &   $\sigma_{\rm rms}$ & $\Delta v$ & Tel. & $t_{\rm int}$ & Ref.\\
\hline%
J0011+1446& 3.4523 & 0.83 & 3.8 & GMRT & 30 & R13\\
B0201+113   &  3.3869 & --- & --- & GMRT & 20 & K14\\
B0335--122  &  3.178  & 1.9 & 6.9  & GMRT & 5.5& K03\\
B0336--017   & 3.0621 & 1.1 & 6.7 & GMRT & 11 & K14\\
B0758+475 & 3.2228  & 0.88 & 3.6 & GMRT & 30 & R13\\
J0816+4823 &    3.3654 & 6.3 & 3.7 & GBT & 6 & C10\\
B1239+376   &  3.411  & 0.8 & 15 & GMRT & 22 &K12\\
B1418--064 & 3.4483 & 11.5 & 3.8 & GMRT & 11& K14\\
J1435+5435   &  3.3032  & 1.5 & 7.1 & GMRT & 6.7 & S12\\
\hline\hline%
\end{tabular}
\label{tab1}
{\small References: K03 -- \citet{kc02}, C10 -- \citet{ctd+09}, K12 -- \citet{kem+12}, S12 -- \citet{sgp+12}, R13 -- \citet{rmgn13}, K14 -- \citet{kps+14}.}
\end{table}
From, this we see that the predicted values are close to those observed and that the fudge factor may even
over-compensate the correction somewhat. Using these conservative estimates, however, we see that sufficiently sensitive
limits, $(1/T_{\rm spin})(d_{\rm abs}/d_{\rm QSO})^2\sim10^{-4} $ K$^{-1}$, are attainable, although only at redshifts
of $z_{\rm abs}\lapp3.5$ (Fig. \ref{sens-current}).

\subsubsection{(Expanded) Very Large Array}

\begin{figure}
\centering \includegraphics[angle=-90,scale=0.55]{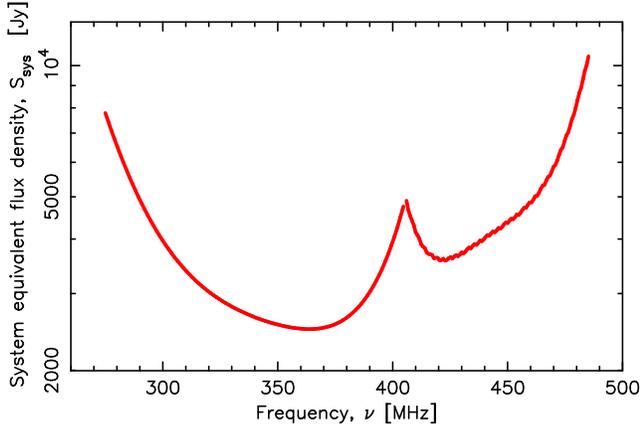}
\caption{Two component (above and below 405 MHz) polynomial fits to the system equivalent flux density of the EVLA P-band.}
\label{VLA}
\end{figure}

The VLA P-band spans 230--470 MHz and so is suitable for 21-cm searches at the redshifts of interest, yielding
408 DR9 DLAs at declinations of $\delta \gapp -30$\dg.  Using the improved sensitivity of the  {\em Expanded
Very Large Array} (EVLA) upgrade (Fig. \ref{VLA}),
with 27 antennas, giving 351 baseline pairs, a correlator efficiency of $\epsilon_{\rm c} = 0.93$, assuming a fudge
factor of 2 in the time estimate\footnote{https://science.nrao.edu/facilities/vla/docs/manuals/propvla/determining}, in dual polarisation,
gives $\sigma_{\rm rms} = 4.7-55$~mJy per 10~\kms, after 10~hours of integration.
As per the GMRT, limits of $(1/T_{\rm spin})(d_{\rm abs}/d_{\rm QSO})^2\sim10^{-4}$~K$^{-1}$ are attainable, but again only at $z_{\rm abs} \lapp3.5$ 
(Fig.~\ref{sens-current}).


\subsection{Prospects with the Square Kilometre Array}

It is therefore apparent that current instruments are unlikely to be able to provide sufficiently sensitive limits to determine
whether $(1/T_{\rm spin})(d_{\rm abs}/d_{\rm QSO})^2$ exhibits the same downturn as the star formation density at high
redshift.  Surveys for 21-cm absorption with the SKA pathfinders, the {\em APERture Tile Array In Focus} (APERTIF), the {\em
  Australian Square Kilometre Array Pathfinder} (ASKAP)  
and MeerKAT ({\em Karoo Array Telescope}) will be limited to $z\lapp0.26$, $z\lapp1.0$ and $z\lapp1.4$, respectively
(see \citealt{mmo+17}).
Therefore, if $(1/T_{\rm spin})(d_{\rm abs}/d_{\rm QSO})^2$ does trace the star formation density, i.e. this is not
ruled out by a number of $z_{\rm abs}\gapp3$ detections with $(1/T_{\rm spin})(d_{\rm abs}/d_{\rm
  QSO})^2\gapp10^{-3}$~K$^{-1}$ (Fig.~\ref{sens-current}), the {\em Square Kilometre Array} will be required to verify
this.  The first phase of the SKA is expected to be complete in 2020, comprising 125\,000 low frequency (50--350 MHz)
antennas, located in Australia, and 300 mid-frequency (350 MHz --14 GHz) dishes, located in South Africa.  Phase-2,
expected around 2028, is planned to comprise one million low frequency antennas and 2000 dishes.
\begin{figure}
\centering \includegraphics[angle=-90,scale=0.55]{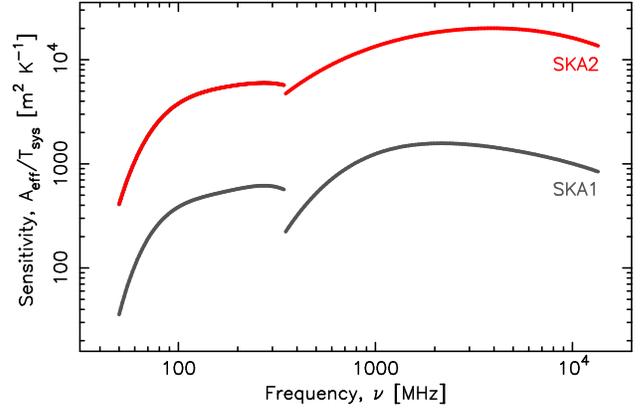}
\caption{Polynomial fits to the point source sensitivity of the SKA \citep{bra17}.}  
\label{SKA-sens}
\end{figure} 
The natural weighted sensitivity for each phase is shown in Fig.~\ref{SKA-sens}. However, again this represents and
ideal and so in Fig. \ref{table1+2} we show the expected sensitivity, at least for the SKA phase-1, where this is 
available  \citep{bra17}.
\begin{figure}
\centering \includegraphics[angle=-90,scale=0.52]{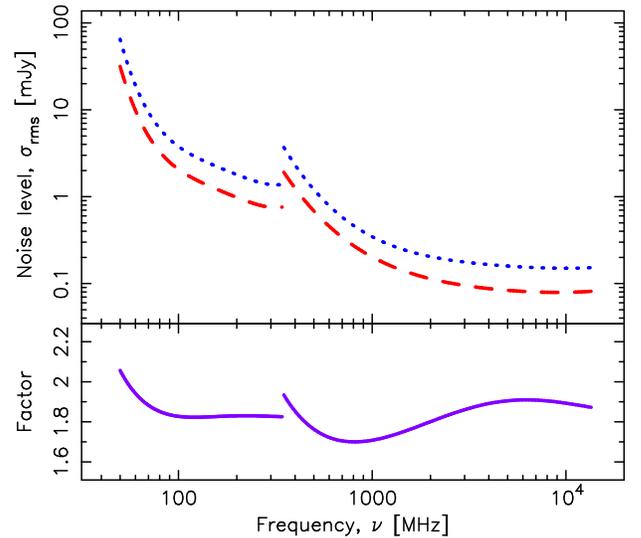}
\caption{The noise level per each 10 \kms\ channel after 1~hour of integration with the SKA1.
The dotted curve shows the natural weighted array sensitivities (Fig. \ref{SKA-sens}) and the broken
curve the expected sensitivities (tables 1 \& 2 of \citealt{bra17}). The bottom panel shows the ratio, i.e.
the fudge factor in the noise level.}
\label{table1+2}
\end{figure} 
From this, we see that the fudge factor is expected to remain close to 2 for both phase-1 low and mid frequency apertures,
which we assume to be the case for the SKA phase-2.

\begin{figure}
\hspace{-1.0em}\includegraphics[angle=-90,scale=0.55]{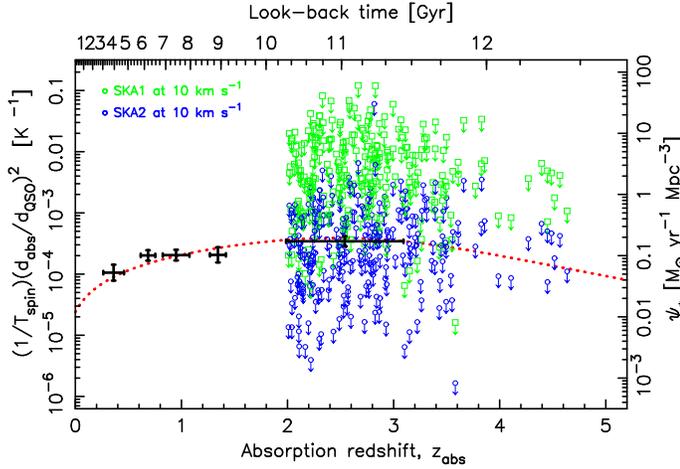}
\caption{As Fig. \ref{sens-current}, but showing the limits reached by the SKA after 10~hours of integration at 10~\kms\
  per channel for sight-lines at $\delta < 29$\dg\ (where 237 absorbers reach elevations of $>30$\dg).}
\label{sens_SKA-dec}
\end{figure}

\begin{figure}
\hspace{-1.0em}\includegraphics[angle=-90,scale=0.55]{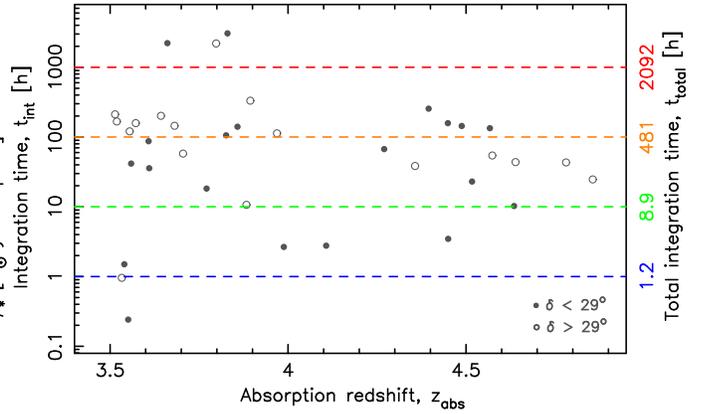}
\caption{The integration times expected by the SKA2-low to obtain a $3\sigma$ detection if $\psi_{*}T_{\rm spin}(d_{\rm
    QSO}/d_{\rm abs})^2 = 500$~\Mo\ yr$^{-1}$ Mpc $^{-3}$ K (e.g. the dotted trace in Fig. \ref{sens_SKA-dec}). The
  filled symbols show the sources at declinations of $\delta < 29$\dg\ and the unfilled those with $\delta > 29$\dg,
  although all of the targets have $\delta\leq41.8$\dg. The right axis shows the total integration time required on the
  basis of the longest single integration for that sight-line. For example, restricting all targets to $t_{\rm int}
  <100$ hours, gives 21 absorbers along 16 sight-lines with $z_{\rm abs} > 3.5$, for  a total observing time of 481 hours.}
\label{time_SKA}
\end{figure}

\begin{figure*}
\centering 
\includegraphics[angle=-90,scale=0.60]{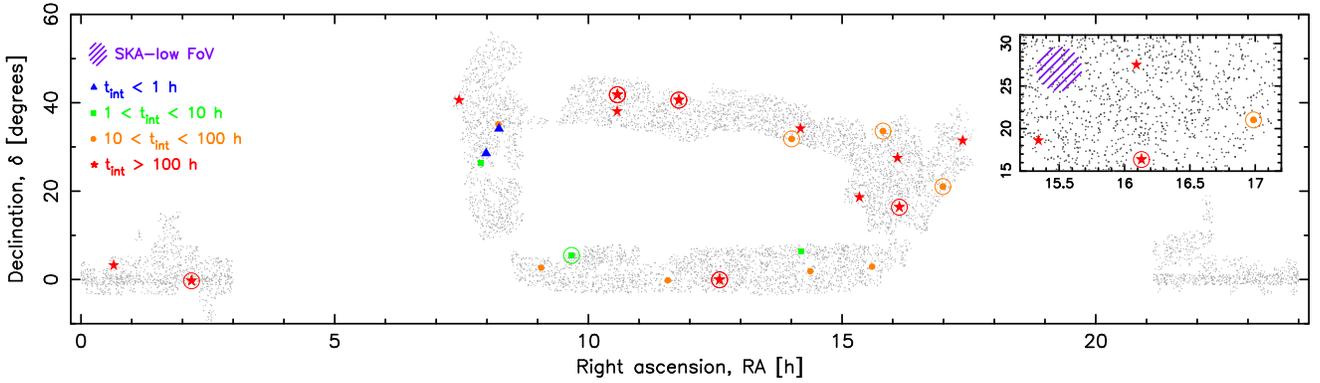}
\caption{The sky distribution of the known radio illuminated $z_{\rm abs} > 3.5$ absorbers. The shapes designate
the maximum required integration time (see Fig.~\ref{time_SKA}) with those circled having more than one 
absorber at $z_{\rm abs} > 3.5$ along the sight-line. The small markers show the positions of the DR9 DLAs and the
hatched region the SKA-low field-of-view (21 deg$^2$, \citealt{dew15}). The inset shows a region of relatively high DLA density.}
\label{RA-dec}
\end{figure*}

Inserting the values for $A_{\rm eff}/T_{\rm sys}$ into Equ. \ref{eq:rad} and scaling by the fudge factor, gives
$\sigma_{\rm rms} = 0.5 - 1.2$ and $0.03-0.06$ mJy per 10 \kms\ for phase-1 and phase-2, respectively. For the
estimated flux densities, these result in $(1/T_{\rm spin})(d_{\rm abs}/d_{\rm QSO})^2\sim10^{-3}$~K$^{-1}$ and
$(1/T_{\rm spin})(d_{\rm abs}/d_{\rm QSO})^2\sim10^{-4}$~K$^{-1}$ at $z_{\rm abs} \gapp3.5$, respectively
(Fig. \ref{sens_SKA-dec}).
We see that these limits approach those required to confirm the downturn in $(1/T_{\rm spin})(d_{\rm
  abs}/d_{\rm QSO})^2$, particularly for the SKA phase-2. Examining this in detail, in Fig.~\ref{time_SKA} we
show the integration times required to reach the necessary sensitivities at $z\gapp3.5$.
From this, we see that most of the absorbers can be searched to sufficiently deep limits within a total of $\sim1000$
hours. Since we used the longest integration required along the sight-lines with multiple absorbers, we expect a number
of observations to be significantly more sensitive than required. Furthermore, if the fraction of cool gas, as traced by
$(1/T_{\rm spin})(d_{\rm abs}/d_{\rm QSO})^2$, does not follow the same steep decline as the star formation density
(Fig.~\ref{SFR-today}) at high redshift, we may expect detections within much shorter integration times.  The wide
SKA-low field-of-view will allow this time to be cut further by observing multiple sight-lines simultaneously
(Fig. \ref{RA-dec}),
which, given the high sky density of DLAs (figure inset), will yield radio flux measurements (rather than estimates, Sect.~\ref{sec:rf}),
at $7"$  resolution \citep{dew15}, and, possibly, unexpected 21-cm absorption.

Looking further, several thousand 21-cm absorbers are expected to be detected with the SKA phase-1 at
$z_{\rm abs} \lapp3$, with no estimate for the numbers at high redshift \citep{msc+15,azdc16}. Even in the
absence of further large DLA catalogues, blind surveys with the SKA are expected to yield a large number of redshifted
absorption systems which are dust reddened (\citealt{wfp+95,cmr+98,cwa+17}) and thus missed by those pre-selected based
upon their optical/UV spectrum. The lack of an optical spectrum does, of course, prevent the nature of the absorber
being determined -- whether it arises in a quiescent galaxy, intervening a more distant continuum source (as per the DLAs), or is
associated with the host of the background continuum source itself. Machine learning techniques do, however, offer the
possibility of determining the nature of the absorber purely from its 21-cm absorption profile \citep{cdda16}. The
other issue with avoiding optical pre-selection is the lack of a Lyman-\AL\ spectrum from which to determine the total neutral
hydrogen column density. A statistical value for use in Equ. \ref{eq:f} may, however, be derived from the cosmological
\HI\ density \citep{cur17a}. 

\section{Discussion and summary}

It is now well established that the evolution of the mass density of neutral hydrogen is in stark disagreement to
that of the star formation density. There is, however, recent compelling evidence that the fraction of cool
gas, as traced through the \HI\ 21-cm absorption strength normalised by the total column density, could trace $\psi_{*}$.
However, the paucity of 21-cm absorption searches at $z_{\rm abs}\gapp3.5$ prevents us from confirming whether
the cold gas fraction follows the same downturn as the star formation density.
In this paper, we examine the possibility of testing this through observations of a large sample of
damped Lyman-\AL\ absorption systems, the SDSS SDSS-III  DR9  catalogue \citep{npc+12}. For each of the 12\,081
$N_{\rm HI}\ge1\times10^{20}$ \scm\ absorption systems, we:
\begin{itemize}
\item Search each sight-line in the NVSS, FIRST and SUMSS for background radio emission. This yields 336 radio-loud
  sight-lines, containing a total of 414 absorbers. For each of these we obtain all of the available radio photometry
  and use this to estimate the flux density at the redshifted 21-cm frequency of  each absorber.
   \item  For each of the current large radio interferometers, we determine which absorbers are redshifted into an
      available band and is visible from the telescope location.
    \item From the instrument specifications, we calculate the sensitivity after a ten hour integration which we 
      combine with the flux density and the neutral hydrogen column density, as well as removing line-of-sight
      geometry effects, to estimate the best expected limit to the spin temperature (degenerate with the ratio of the
      emitter--absorption extents) obtainable for each absorber.
\end{itemize}
From this, we find that none of the current instruments are of sufficient sensitivity to place useful limits on
$(1/T_{\rm spin})(d_{\rm abs}/d_{\rm QSO})^2$ at $z_{\rm abs}\gapp3.5$, although both the upgraded (u)GMRT and (E)VLA
may be sufficiently sensitive at $3\lapp z_{\rm abs}\lapp3.5$.  This, of course, does not preclude the possibility
that a number of detections at these redshifts could rule out the hypothesis that $\psi_{*}$ is traced by
$(1/T_{\rm spin})(d_{\rm abs}/d_{\rm QSO})^2$, a possibility which can only be addressed through observation.

With the SKA, the required sensitivity is achievable at $z_{\rm abs}\gapp3.5$), but only for a small number ($\approx10$) of absorbers.
This is due to the sensitivity function of the SDSS, exhibiting a steep decline in the number of sight-lines at $z_{\rm
  abs}\gapp3$ ($\lambda_{\rm obs}\gapp5000$~\AA, \citealt{npls09,npc+12}), above which the Lyman-\AL\ transition is
shifted out of the optical and into the near-infrared band (at $z_{\rm abs}\gapp5.5$). If our derived radio fluxes are
representative at these redshifts, in addition to new DLAs found through further optical surveys over the next
decade, we expect an increased number of absorbers to be detected through the wide instantaneous bandwidth and
field-of-view of the SKA. The detection of new absorption systems in radio surveys would not be subject to the same dust
obscuration as their optical counterparts. In the absence of an optical spectrum, it may be possible to determine
whether the absorption is intervening or associated, through machine learning techniques \citep{cdda16} and apply a
statistical column density in order to yield a statistical $(1/T_{\rm spin})(d_{\rm abs}/d_{\rm QSO})^2$ \citep{cur17a}.

Radio selection may also uncover intervening absorbers rich in molecular gas: Although molecular absorption has been
detected in 26 DLAs, through \MOLH\ vibrational transitions redshifted into the optical band at $z_{\rm abs}\gapp1.7$
(compiled in \citealt{sgp+10} with the addition of \citealt{rbql03,fln+11,gnp+12,sgp+12,nsr+15,nkb+16,bnr+17}),
extensive millimetre-wave observations have yet to detect absorption from any rotational transition
(e.g. \citealt{cmpw03}). Since the DLAs in which H$_2$ has been detected have molecular fractions ${\cal
  F}\equiv{2N_{\rm H_2}}/{(2N_{\rm H_2}+N_{\rm HI})}\sim10^{-7}-0.3$ and optical -- near-infrared colours of $V-K\lapp4$
(\citealt{cwc+11}), compared to the five known redshifted radio-band absorbers, where ${\cal F}\approx0.7-1.0$ and
$V-K\geq4.80$ \citep{cwm+06}, we suspect that the selection of optically bright objects selects against dusty
environments, which are more likely to harbour molecules in abundance. Thus, radio selected surveys offer the
possibility of finally detecting dust obscured DLAs which have similarly high molecular fractions (${\cal
  F}\sim1$). Comparison of the atomic and molecular line strengths will provide an invaluable probe of the conditions in
the highest redshift galaxies.  Furthermore, comparison of the relative shifts of the atomic and molecular transitions
in the radio-band offer a measure of the fundamental constants of nature at large-look back times to much greater
precision than optical spectroscopy (e.g. \citealt{mwf03,tmw+06}), making the SKA ideal in resolving this contentious
issue \citep{cdk04}.

\noindent{\bf Acknowledgements}
 I wish to thank the referee for their very helpful comments, as well as James Allison and Randall Wayth for their help
  with the MWA specifications, Raymond Oonk, Vanessa Moss and Antonis Polatidis for the LOFAR specifications and Robert
  Braun for the SKA specifications.  This research has made use of the NASA/IPAC Extragalactic Database (NED) which is
  operated by the Jet Propulsion Laboratory, California Institute of Technology, under contract with the National
  Aeronautics and Space Administration and NASA's Astrophysics Data System Bibliographic Service.  


\label{lastpage}

\end{document}